\documentclass[11pt,twoside]{article}
\usepackage{asp2004}
\usepackage{graphicx}
\def\edcomment#1{\iffalse\marginpar{\raggedright\sl#1\/}\else\relax\fi}
\reversemarginpar

\begin{document}
\title{Dynamic Tides in Close Binaries}
\author{B. Willems}
\affil{Northwestern University, Department of Physics and Astronomy,
  2145 Sheridan Road, Evanston, IL 60208, USA}

\begin{abstract}
The basic theory of dynamic tides in close binaries is
reviewed. Particular attention is paid to resonances between dynamic
tides and free oscillation modes and to the role of the
apsidal-motion rate in probing the internal 
structure of binary components. The discussed effects are
generally applicable to stars across the entire Hertzsprung-Russell
diagram, including the binary OB-stars discussed at this meeting.  
\end{abstract}
\thispagestyle{plain}


\section{Introduction}

In close binaries, each star is distorted by the tidal action exerted
by its companion. The tides give rise to a variety of dynamical
effects from the resonant excitation of free oscillation modes to the
secular evolution of the orbital elements. The effects are richest
when the binary is eccentric and the component stars are rotating
asynchronously with respect to the orbital motion. The tidal force is
then time-dependent and the tides are known as {\em dynamic
tides}. Due to viscous and dissipative effects, the tidal distortion
lags behind the position of the companion, creating a torque which
tends to circularize the orbit and synchronize the component
stars. The time scales of circularization and synchronization depend
strongly on the initial binary parameters and on the physical
processes responsible for the dissipation of energy. Once
circularization and synchronization are achieved, the tidal action is
static with respect to a frame of reference corotating with the
stars\footnote{In order for this equilibrium state to set in, the
stars' equatorial planes must also coincide with the orbital
plane. Throughout this paper, we will always assume this to be the
case.}. These tides are known as {\em equilibrium tides}.

The dependence of tidal effects on processes taking place in the
stellar interior is a fortuitous circumstance that allows tides to be
used to probe physics hidden below the stellar surface. In order
to exploit this probing potential, a thorough theoretical
understanding is, however, essential. In the following sections, we
therefore discuss some basic aspects of the theory of dynamic tides in
close binaries. Highlighted applications are the excitation of
oscillation modes by resonant dynamic tides and the role of the
apsidal-motion rate in studying components of eccentric binaries. We
conclude with a brief discussion of future prospects and applications
such as the study of the formation of compact objects, the detection
of gravitational waves by the NASA/ESA cornerstone mission ({\em
LISA}), and the contribution of large scale photometric variability
and exoplanet transit surveys to the study of tides in close binaries.

\section{The tide-generating potential}

We consider a close binary system of stars revolving around each other
under the influence of their mutual gravitational force. We denote the
orbital period by $P_{\rm orb}$, the semi-major axis by $a$, and the
eccentricity by $e$. The first star, with mass $M_1$ and radius $R_1$,
is assumed to rotate uniformly with an angular velocity $\Omega_{\rm
rot}$ perpendicular to the orbital plane. The companion star, with
mass $M_2$, is treated as a point mass.

The tidal force exerted by the companion is derived from the
tide-generating potential $\varepsilon_T\,W(\vec{r},t)$, where
$\varepsilon_T = (R_1/a)^3 (M_2/M_1)$. In binaries with circular
orbits, the parameter $\varepsilon_T$ corresponds to the ratio of the
tidal force to the gravity at the star's equator, which is
approximately equal to the height of the tidal bulge raised on the
primary by the secondary. 

The tidal distortion of the primary is most
conveniently studied by expanding the tide-generating potential in
terms of unnormalized spherical harmonics $Y_\ell^m(\theta,\phi)$ and
in Fourier series in terms of multiples of the mean motion
$n=2\pi/P_{\rm orb}$:
\begin{eqnarray}
\lefteqn{ \varepsilon_T\, W \left( \vec{r},t \right) = -
  \varepsilon_T\, {{G\, M_1} \over R_1}\,
  \sum_{\ell=2}^4 \sum_{m=-\ell}^\ell \sum_{k=-\infty}^\infty
  c_{\ell,m,k}\, \left( {r \over R_1} \right)^\ell
  } \nonumber \\
& & \times\,
  Y_\ell^m (\theta,\phi)\, \exp \left[ {\rm i}
  \left( \sigma_T\, t - k\, n\, \tau \right) \right]
  \hspace{3.5cm} \label{pot}
\end{eqnarray}
(e.g. Zahn 1970, Polfliet \& Smeyers 1990). Here, $G$ is the
Newtonian gravitational constant, $(r,\theta,\phi)$ is a system of
spherical coordinates with respect to a frame of reference that is
co-rotating with the star, $\sigma_T = k\, n + m\, \Omega$ are forcing
angular frequencies with respect to the co-rotating frame of
reference, and $\tau$ is a time of periastron passage. The factors
$c_{\ell,m,k}$ are Fourier coefficients defined as
\begin{eqnarray}
\lefteqn{c_{\ell,m,k} = \displaystyle
  {{(\ell-|m|)!} \over {(\ell+|m|)!}}\, P_\ell^{|m|}(0)
  \left({R_1\over a}\right)^{\ell-2}
  {1\over {\left({1 - e^2}\right)^{\ell - 1/2}}} } \nonumber \\
 & & \times {1\over \pi} {\int_0^\pi (1 + e\, \cos v)^{\ell-1}\,
  \cos (k\, M + m\, v)\, dv}, \label{clmk}
\end{eqnarray}
where $P_\ell^{|m|}(x)$ is an associated Legendre polynomial of the
first kind, and $M$ and $v$ are the mean and true anomaly of the
companion in its relative orbit, respectively. The coefficients
$c_{\ell,m,k}$ obey property of symmetry $c_{\ell,-m,-k} =
c_{\ell,m,k}$ and are equal to zero for odd values of $\ell+|m|$. The
coefficients $c_{\ell,m,0}$ are equal to zero for $|m| > \ell-1$.

\begin{figure}
\resizebox{6.35cm}{!}{\includegraphics{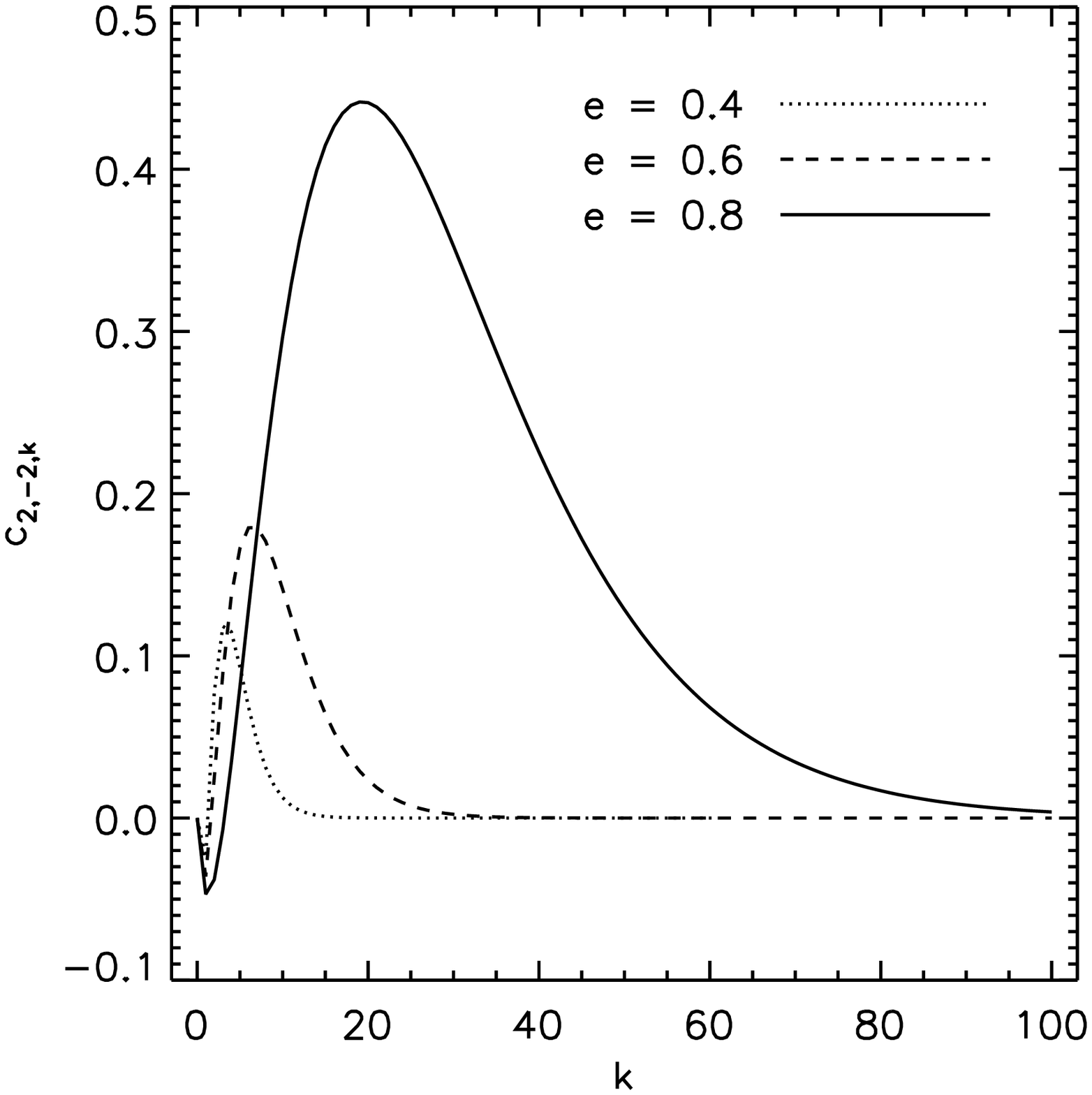}}
\resizebox{6.35cm}{!}{\includegraphics{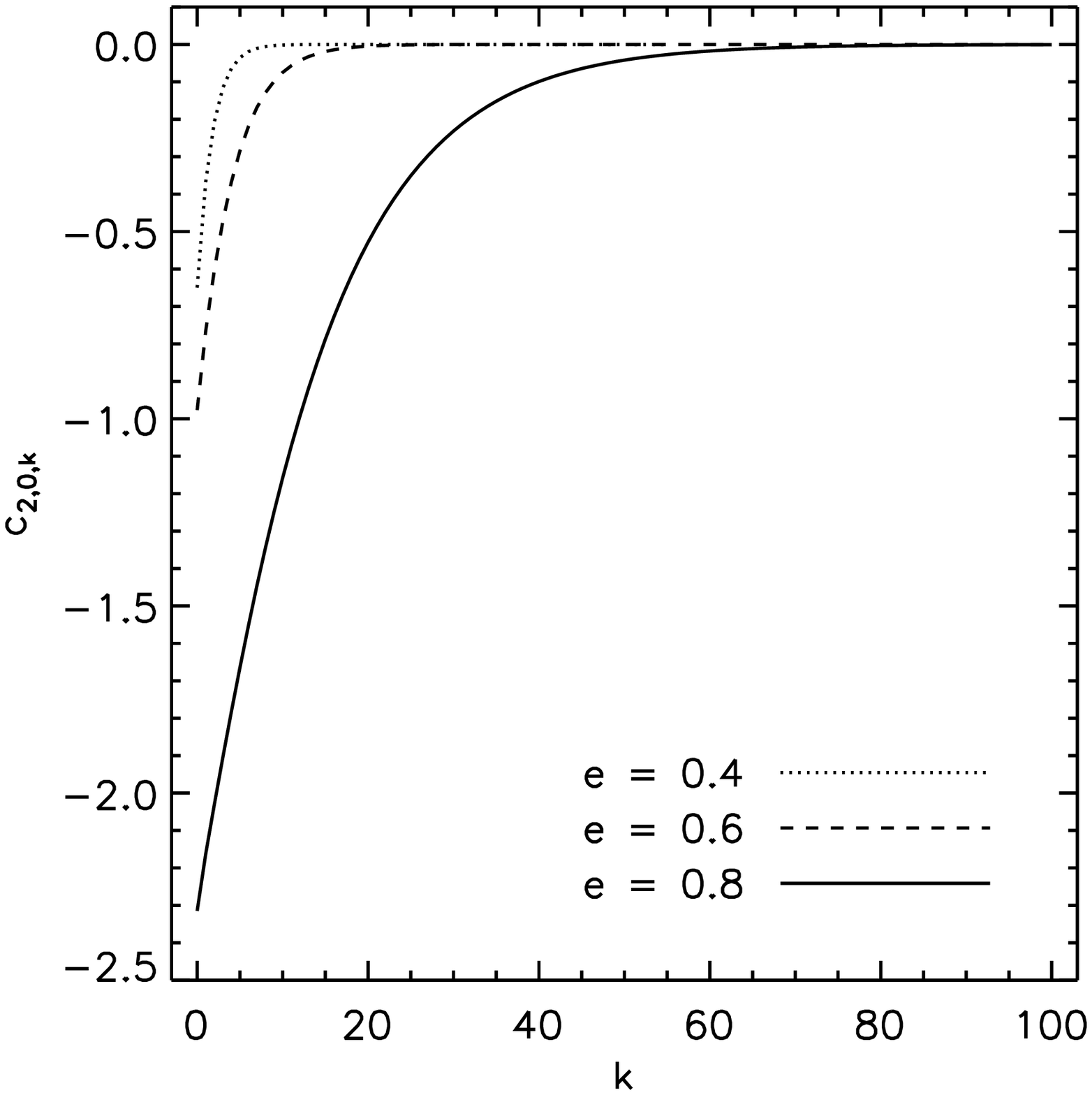}}
\caption{The Fourier coefficients $c_{2,-2,k}$ (left) and $c_{2,0,k}$
  (right) as functions of $k$, for $e=0.4$, $0.6$, and $0.8$. The
coefficients $c_{2,2,k}$ tend to be several orders of magnitude smaller
than the coefficients $c_{2,-2,k}$ and $c_{2,0,k}$, and are therefore
less important in the expansion of the tide-generating potential (see
  Willems 2003 for details).}   
\label{fclmk}
\end{figure}

The dominant terms in Expansion (\ref{pot}) of the tide-generating
potential are the terms associated with $\ell=2$. The associated
Fourier coefficients are independent of $R_1/a$ and are different from
zero only for $m=0$ and $m=\pm 2$. The variations of the coefficients
$c_{2,-2,k}$ and $c_{2,0,k}$ as functions of $k$ are shown in
Fig.~\ref{fclmk} for three different orbital eccentricities $e$.
For a given orbital eccentricity and sufficiently high $k$-values, the
coefficients decrease in absolute value with increasing values of $k$,
but the decrease is slower for higher orbital eccentricities. The
number of non-trivially contributing terms in the expansion of the
tide-generating potential therefore increases with increasing values
of $e$. For a given orbital eccentricity, the coefficients $c_{2,-2,k}$ 
furthermore reach a maximum for $k$-values for which $k\,n \approx
\Omega_{\rm peri}$, where $\Omega_{\rm peri}$ is the orbital angular
velocity of the companion at the periastron of its relative
orbit. This is in line with our intuitive expectations that the tidal
forcing in eccentric orbits is strongest when the stars pass through
the periastron of their relative orbit.

\section{Resonant dynamic tides}

The Fourier expansion of the tide-generating potential induces an
infinite number of forcing frequencies in the tidally distorted
star. Non-zero forcing frequencies give rise to
  dynamic tides, while zero forcing frequencies give rise
  to static tides. When one of the forcing frequencies is close to the
eigenfrequency of a free oscillation mode, a resonance occurs and the
mode involved in the resonance is excited by the tidal action exerted
by the companion. The possibility of resonances between dynamic tides
and free oscillation modes is particularly relevant for the excitation
of free oscillation modes $g^+$ since their eigenfrequencies are most
likely to be in the range of the forcing frequencies induced in
components of close binaries.

In what follows, we consider the effects of a resonance between a
dynamic tide associated with the spherical harmonic
$Y_\ell^m(\theta,\phi)$ and the forcing angular frequency
$\sigma_T=k\,n+m\,\Omega$, and a free oscillation mode of radial order
$N$ with eigenfrequency $\sigma_{\ell,N}$ and coefficient of
vibrational stability $\kappa_{\ell,N}$\footnote{From the
orthogonality properties of the spherical harmonics, it follows that
tides associated with the spherical harmonic $Y_\ell^m (\theta,\phi)$
can only excite oscillation modes associated with the same $Y_\ell^m
(\theta,\phi)$ (see Smeyers et al. 1998 for details).}. At the
lowest order of approximation, the tide gives rise to a displacement
field
\begin{equation}
\vec{\xi}_{\rm res} \left( \vec{r},t \right) =  
  {{\varepsilon_T\, c_{\ell,m,k}} \over 2}
  {{\sigma_{\ell,N}\, 
  {\cal Q}_{\ell,N}\, \vec{\xi}_{\ell,N}\left( \vec{r} \right) } \over 
  {\left[ \left( \sigma_{\ell,N} - \sigma_T \right)^2 
  + \kappa_{\ell,N}^2 \right]^{1/2}}}\, 
  \exp \left[ {\rm i}
  \left( \sigma_T\, t - k\, n\, \tau + \psi_{\rm res} \right) 
  \right], \label{xit}
\end{equation}
where $\vec{\xi}_{\ell,N}$ is the vector of the Lagrangian displacement
of the oscillation mode involved in the resonance,  
\begin{equation}
{\cal Q}_{\ell,N} = {{G\,M_1} \over R_1}\, 
   {{\displaystyle \int_{M_1} \vec{\xi}_{\ell,N} 
   \left( \vec{r}\, \right) 
   \cdot \nabla \left[ \left( r/R_1 \right)^\ell 
   Y_\ell^m (\theta,\phi) \right]\, dm} 
   \over {\displaystyle \sigma_{\ell,N}^2\, \int_{M_1}  
   \left| \vec{\xi}_{\ell,N} \left( 
   \vec{r}\, \right) \right|^2 dm}} \label{Qln}
\end{equation}
is the so-called resonance coefficient (Zahn 1970) or overlap integral
(Press \& Teukolsky 1977), and  
\begin{equation}
\psi_{\rm res} = - \arctan {\kappa_{\ell,N} \over 
   {\sigma_{\ell,N} - \sigma_T}} \label{psi}
\end{equation}
is the phase shift between the tidal displacement field and the
resonant term in the expansion of the tide-generating potential
(Willems et al. 2003).   

\begin{figure}
\resizebox{6.35cm}{!}{\includegraphics{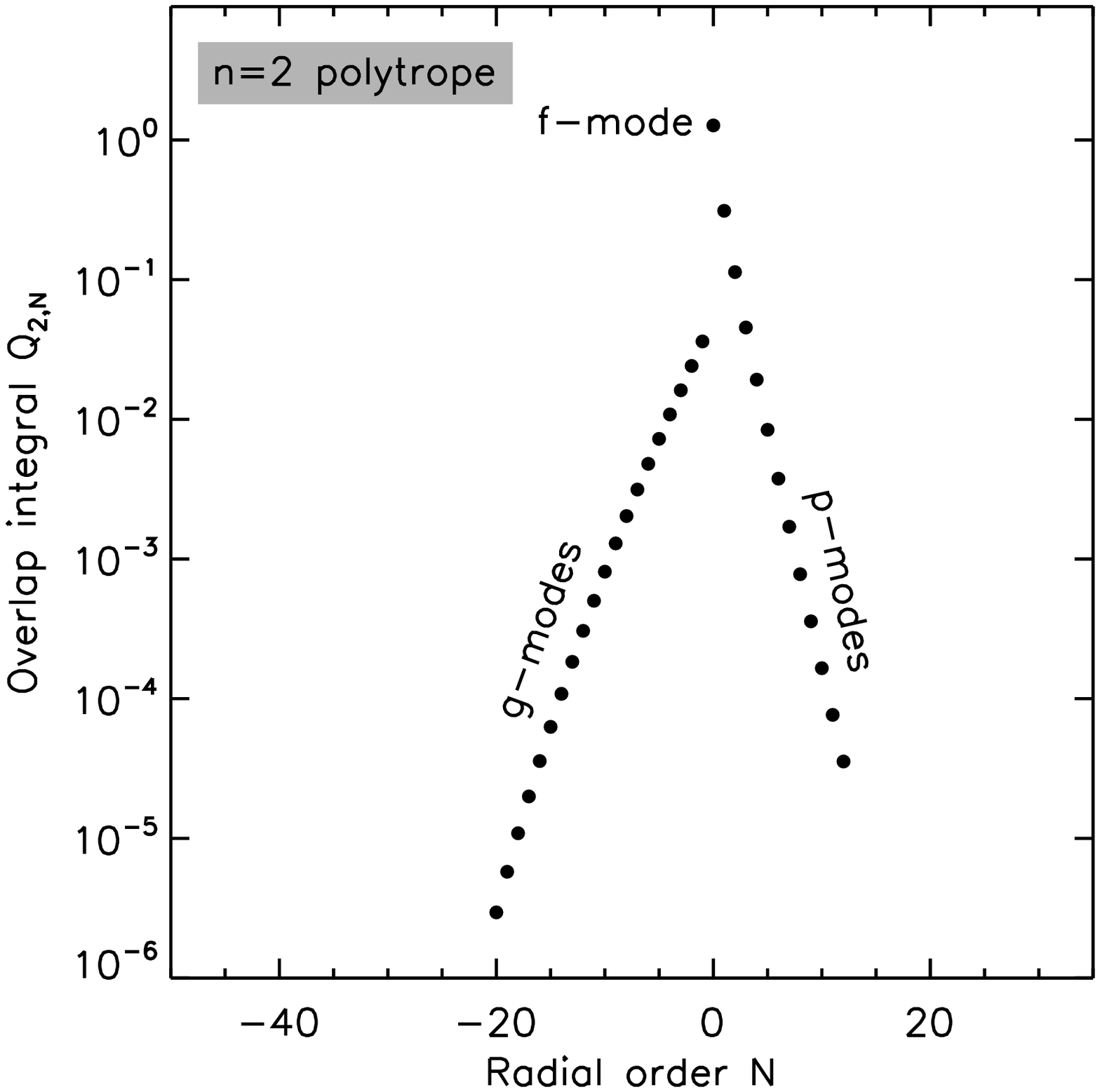}}
\resizebox{6.35cm}{!}{\includegraphics{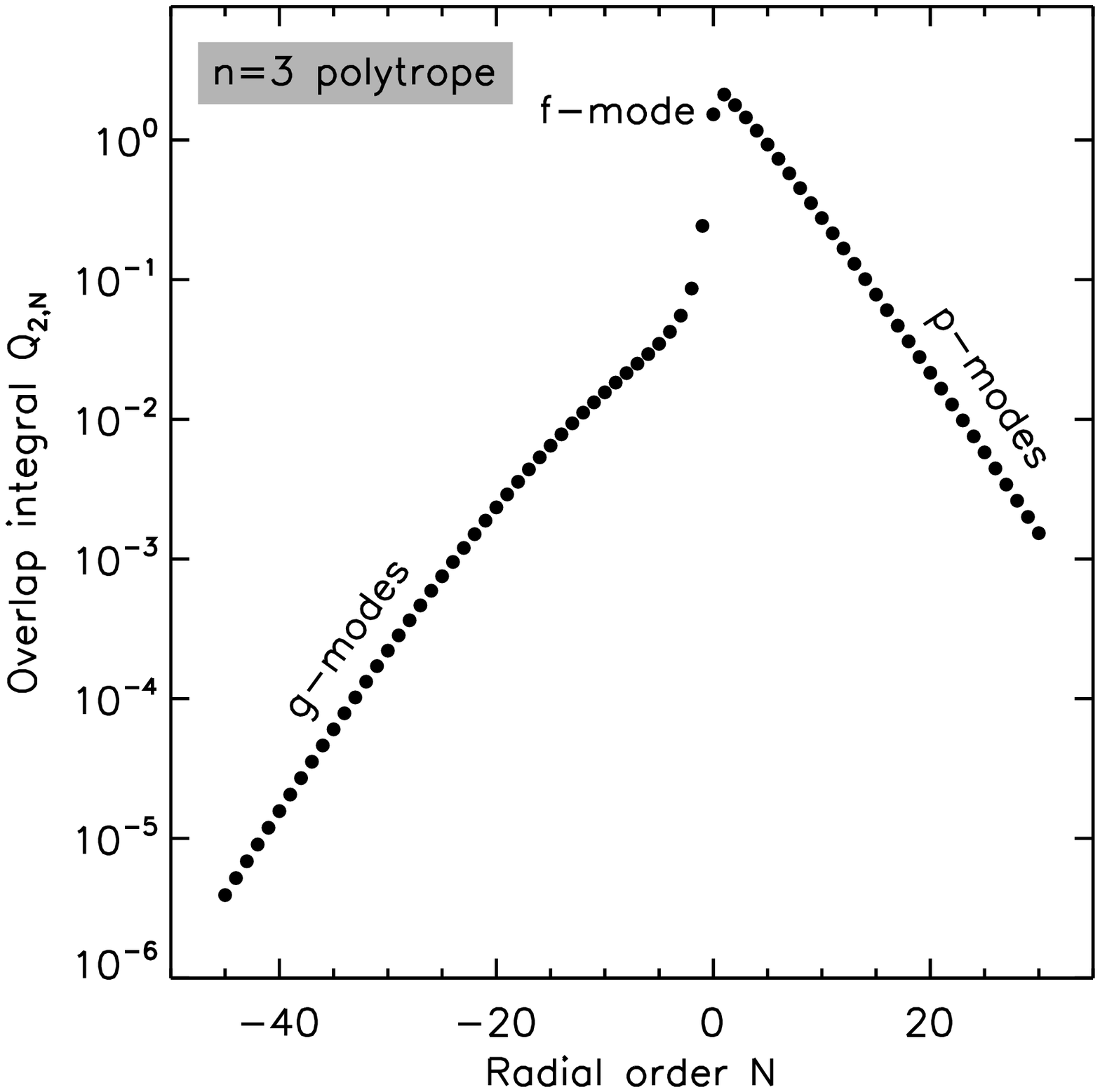}}
\caption{The overlap integrals ${\cal Q}_{2,N}$ for the
  polytropic stellar models with indices $n=2$ (left) and $n=3$
  (right). $g^+$-modes are denoted by negative radial orders,
  $p$-modes by positive radial orders, and the $f$-mode by radial
  order 0. All overlap integrals are determined by normalizing the
  radial component of the Lagrangian displacement of the resonant 
  modes at the star's surface to unity.}
\label{fQln}
\end{figure}

The overlap integral ${\cal Q}_{\ell,N}$ is proportional to the work
done by the tidal force through the resonantly excited mode. For
main-sequence stars, ${\cal Q}_{\ell,N}$ generally decreases rapidly
with increasing values of $N$, so that resonant excitation tends to be
easier for lower order modes than for higher order modes. The cut-off
radial order above which resonant excitation becomes less feasible
depends strongly on the internal structure of the star. This is
illustrated in Fig.~\ref{fQln} for the $\ell=2$ overlap integrals in
two polytropic models with different degrees of central
condensation. For highly condensed stellar models, the behavior of the
overlap integral becomes considerably more erratic than the smooth
trends displayed in Fig.~\ref{fQln}, although the decrease of ${\cal
Q}_{\ell,N}$ with increasing $N$ is generally conserved for
sufficiently high-order modes. Evolved stellar models with sharp
transition zones between layers of different chemical compositions
however show a considerably more complicated pattern of overlap
integrals. Fontaine et al. (2003), for instance, have shown that mode
trapping and confinement effects of $g^+$-modes in subdwarf B stars,
lead to increasing ${\cal Q}_{\ell,N}$-values with increasing radial
order $N$. These stars are therefore extremely susceptible to
resonances between dynamic tides and free oscillation modes, and are
potentially some of the most promising sources to look for forced
oscillations in close binaries.

Resonant dynamic tides can furthermore significantly accelerate the
secular evolution of the binary's semi-major axis and the star's
rotational angular velocity (Savonije \& Papaloizou 1983, 1984;
Willems et al. 2003). Since the forcing angular frequencies $\sigma_T$
depend on $n$ and $\Omega_{\rm rot}$, one would therefore expect
resonances to be fairly short-lived as the binary evolves rapidly
through and away from them. Witte \& Savonije (1999, 2001), however,
have shown that when stellar and orbital evolution are both taken into
account, the changes in the forcing frequencies and the
eigenfrequencies may compensate one another, locking the binary in a
resonance for a prolonged period of time. Such resonance lockings
greatly improve the detectability of resonantly excited oscillation
modes through the enhanced variability of the star's surface
properties or through the enhanced secular evolution of the binary
and its component stars.

\begin{table}
\caption{Examples of stars oscillating with observed frequencies
  $\sigma_{\rm obs}$ equal to integer multiples of the orbital
  frequency $n$.}
\label{stars} 
\smallskip
\begin{center}
\begin{tabular}{lcccccc}
\tableline
\noalign{\smallskip}
Name & $M_1$ & $M_2$ & $P_{\rm orb}$ & $e$ & $\sigma_{\rm obs}/n$ &
Refs. \\
\noalign{\smallskip}
\tableline
\noalign{\smallskip}
 HD\,177863 & $3.5M_\odot$ & $1.0\!-\!2.0M_\odot$ & 11.9\,d  & 0.60 &
 10 & 1, 2, 3 \\
 HD\,209295 & $1.8M_\odot$ & $0.6\!-\!1.5M_\odot$ &  3.11\,d & 0.35 &
 3, 5, 7, 8, 9 & 4 \\ 
 HD\,77581  & $23\!-\!29M_\odot$ & $1.8\!-\!2.4M_\odot$ & 8.96\,d & 0.09 &
 1, 4 & 5 \\
\noalign{\smallskip}
\tableline
\end{tabular}
\end{center}
{\small References: (1) De Cat et al. 2000; (2) De Cat 2001,
  (3) Willems \& Aerts 2002; (4) Handler et al. 2002; (5) Quaintrell
  et al. 2003.}
\end{table}

\begin{figure}
\begin{center}
\resizebox{6.35cm}{!}{\includegraphics{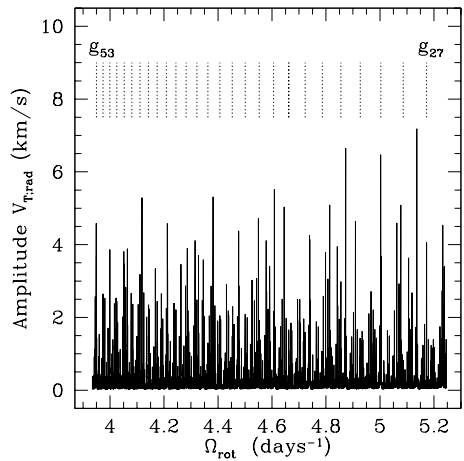}}
\end{center}
\caption{Theoretically predicted amplitude of the $\ell=2$ tidally
  induced radial-velocity variations in HD\,177863 as a function of
  the unknown rotational angular velocity $\Omega_{\rm rot}$. The
  orbital inclination was set to $i=35^\circ$ and the companion mass
  to $M_2=2\,M_\odot$.  The dashed vertical lines indicate resonances
  for which $\sigma_{\rm obs} \approx 10\,n$, corresponding to the
  observed commensurability of frequencies (see
  Table~\ref{stars}). Details of the calculation can be found in
  Willems \& Aerts (2002).}
\label{hd177863}
\end{figure}

Despite the possibility of resonance lockings, firm evidence for the
presence of resonantly excited oscillation modes in components of
close binaries remains scarce. The most promising candidates so far
are the multi-periodic oscillators HD\,177863, HD\,209295, and
HD\,77581 listed in Table~\ref{stars}. In all three cases, the primary
exhibits pulsations with frequencies equal to an integer multiple of
the orbital frequency, strongly suggesting the possibility of
resonantly excited oscillation modes (note that in the observer's
non-rotating frame of reference the condition for a mode with an
observed frequency $\sigma_{\rm obs}$ to be resonantly excited is
$\sigma_{\rm obs} \approx k\,n$). Additional support for the resonant
excitation of these modes is hard to obtain due to the lack of
accurately known rotational angular velocities. Willems \& Aerts
(2002), e.g., have shown that in the case of HD\,177863, the errors on
the rotational angular velocity introduce uncertainties in the forcing
frequencies that are larger than the typical separation between two
successive resonances. Theoretically one therefore finds a whole range
of possible resonances (see Fig.~\ref{hd177863}), which severely
hinders any definite identification of the resonantly excited mode.

\section{Apsidal motion}

The tidal distortion of close binary components perturbs the spherical
symmetry of the external gravitational field. This perturbation in
turn gives rise to time-dependent variations of the orbital elements,
the magnitude of which depends on the internal structure of the
stars. The motion of the stars in the perturbed gravitational field
can therefore be used to probe the interior of close binary
components.

Particularly useful for this purpose is the rate of secular change of
the longitude of the periastron $\varpi$. The phenomenon is periodic
on time scales of the order of $10$--$10^6$\,years, and is caused by a
combination of the tidal and rotational distortions of the binary
components as well as general relativistic effects. Here, we focus on
the contribution of tides to the rate of secular apsidal
motion. Elaborate discussions and references on the contribution of
rotational and general relativistic effects can be found in Gim\'enez
(1985), Claret \& Gim\'enez (1993), and Claret \& Willems (2002).

The most commonly used formula for the rate of secular apsidal motion
is due to Cowling (1938) and Sterne (1939). These authors derived the
contribution of the tidal distortion to the apsidal-motion rate under
the assumption that the orbital and rotational periods are long in
comparison to the periods of the free oscillation modes of the
component stars. The tidal distortion can then be approximated as
static and the resulting apsidal-motion rate due to the dominant
$\ell=2$ tides is given by
\begin{equation}
\left( {{d\varpi} \over {dt}} \right)_{\rm tides} = 
  \left( {R_1 \over a} \right)^5 {M_2 \over M_1}\, 
  {{2\,\pi} \over P_{\rm orb}}\, k_2\, 15\, 
  f\!\left( e^2 \right), \label{sterne}
\end{equation}
where 
\begin{equation}
f\!\left(e^2\right) = \left({1-e^2}\right)^{-5}\,
  \left({1 + {3\over 2}\, e^2 + {1\over 8}\, e^4 }\right). 
\end{equation}
The constant $k_2$, which is known as the apsidal-motion constant,
depends on the internal structure of the star and measures the extent
to which mass is concentrated towards the stellar center. The constant
takes the value $k_2=0$ in the case of a point mass and the value
$k_2=0.75$ in the case of the equilibrium sphere with uniform mass
density. For main-sequence stars, $k_2$ is typically of the order of
$10^{-3}-10^{-2}$.

The fairly straightforward dependence on the internal density profile
and the independence on less well constrained dissipative effects such
as radiative and convective damping are the main ingredients that make
the apsidal-motion rate a simple but powerful tool to study
stellar physics. Comparisons between theoretically and
observationally derived apsidal-motion rates (and thus between
theoretical and observational $k_2$-values) go back to Schwarzschild
(1958) and continue to be updated as new observations and improved
stellar models become available (e.g. Claret \& Gim\'enez 1993, Claret
\& Willems 2002). Much of the simplicity of Eq.~(\ref{sterne}) is,
however, related to the assumption that the orbital and
rotational periods are long in comparison to the periods of the free
oscillation modes of the component stars. When this assumption is
violated, dynamical effects become important and the time-dependent
response of the star to the tidal forcing must be taken into account
(Smeyers \& Willems 2001). 

The rate of secular apsidal motion accounting for the effects of
dynamic tides can be cast in a form similar Eq.~(\ref{sterne}),
provided that the apsidal-motion constant $k_2$ is replaced by a
generalized apsidal-motion constant $k_{\rm 2,dyn}$. The latter
depends on the orbital and rotational periods as well as on the
orbital eccentricity, and can be interpreted as a weighted mean of the
response of the star to the forcing frequencies $\sigma_T$
(Claret \& Willems 2002). The constant $k_{2,{\rm dyn}}$ can be
negative as well as positive, so that periastron recessions can occur
in addition to periastron advances. These periastron recessions,
however, usually only arise for close resonances in which the forcing
frequency of the resonant dynamic tide is slightly larger than the
eigenfrequency of the oscillation mode involved in the resonance.

\begin{figure}
\resizebox{6.35cm}{!}{\includegraphics{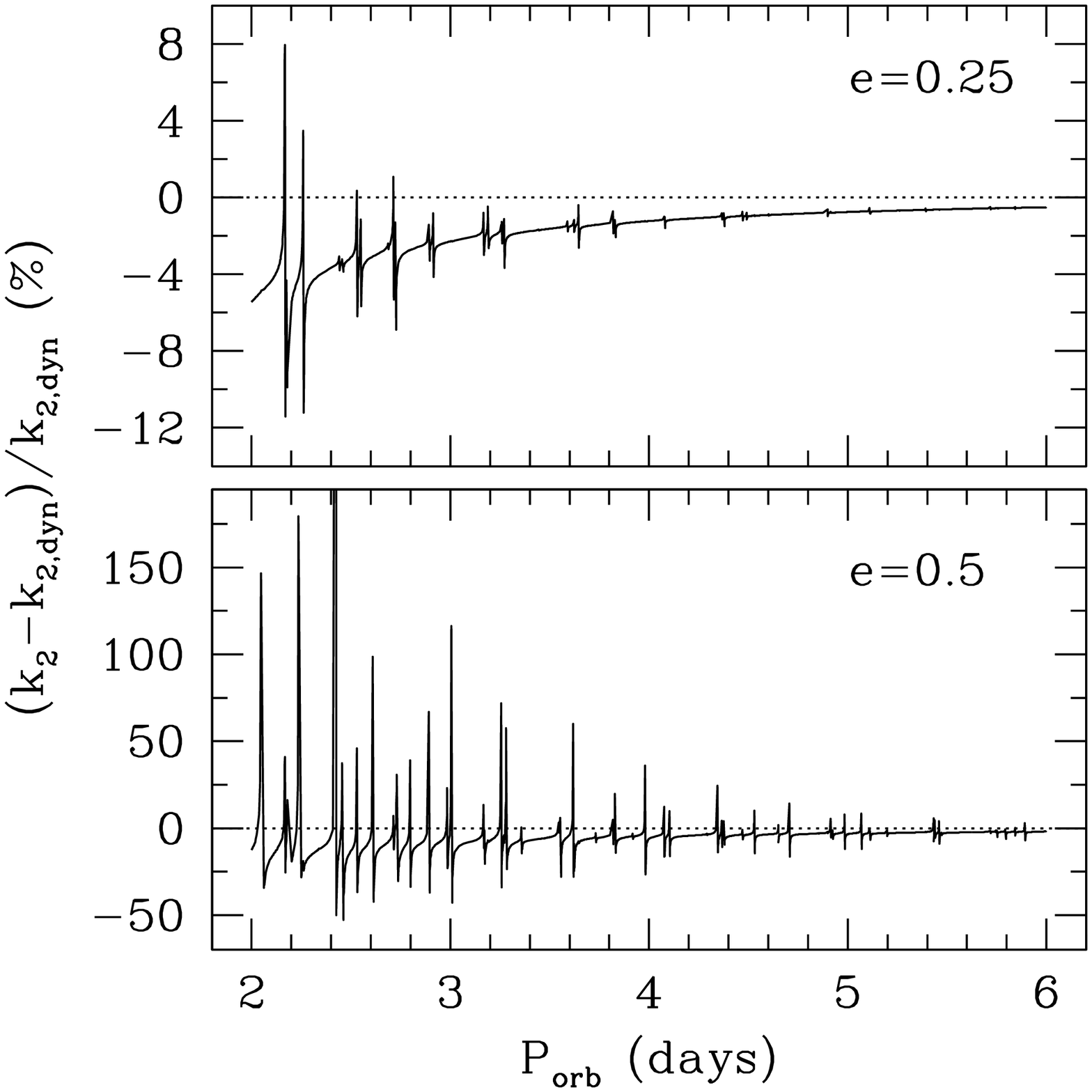}}
\resizebox{6.35cm}{!}{\includegraphics{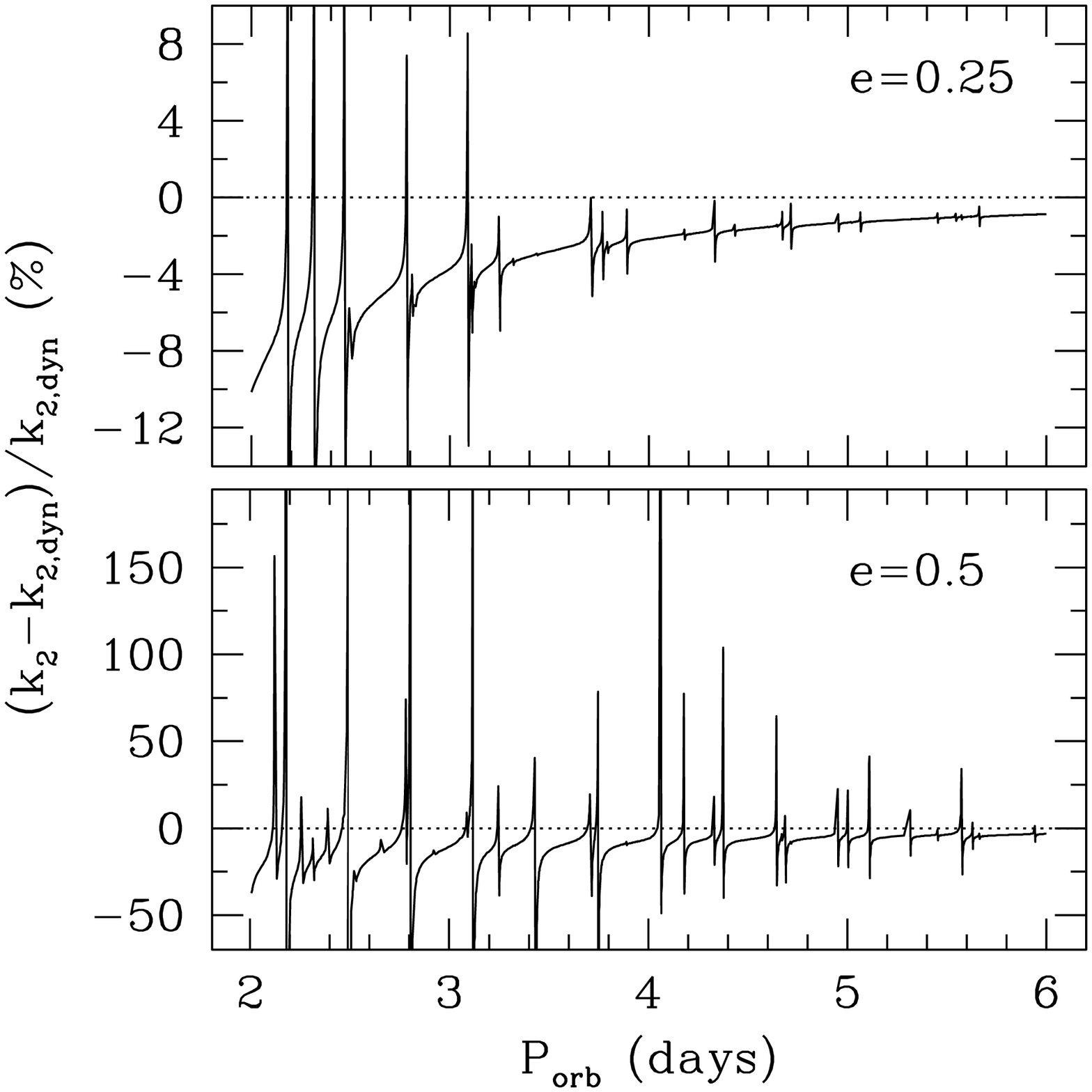}}
\caption{Relative differences between the classical apsidal-motion
  constant $k_2$ and the generalized apsidal-motion constant $k_{\rm
  2,dyn}$ accounting for the effects of dynamic tides, for
  $10\,M_\odot$ (left) and $20\,M_\odot$ (right) ZAMS stellar models
  and orbital eccentricities $e=0.25$ and $e= 0.5$. The rotational
  angular velocity was assumed to be small and set equal to
  $\Omega_{\rm rot} = 0.01\,n$. Details of the calculation can be
  found in Smeyers \& Willems (2001).}
\label{aps}
\end{figure}

The effects of dynamic tides on the rate of secular apsidal motion for
$10\,M_\odot$ and $20\,M_\odot$ ZAMS stars are illustrated in
Fig.~\ref{aps} where the relative differences between $k_2$ and
$k_{\rm 2,dyn}$ are displayed as a function of the orbital period for
two different orbital eccentricities. The relative differences are
mostly negative, so that for binaries with shorter orbital periods
Eq.~(\ref{sterne}) yields somewhat too small values for the rate of
secular apsidal motion, and thus somewhat too long apsidal-motion
periods. The peaks observed at shorter orbital periods are caused by
resonances of dynamic tides with free oscillation modes $g^+$ of the
tidally distorted star. These peaks are superposed on a basic curve
which represents the systematic deviations caused by the increasing
role of the stellar compressibility at shorter orbital periods.

The systematic deviations due to the stellar compressibility increase
with increasing mass of the ZAMS star. This behaviour is related to
the decrease in the amount of central condensation and the associated
increase in the impact of tides with increasing stellar mass. The
systematic deviations also become larger with increasing orbital
eccentricity due to the larger number of higher-frequency tides
contributing to the tide-generating potential.  The number of resonant
dynamic tides on the other hand tends to decrease with increasing
stellar mass. The reason for this is that ZAMS stars with higher
masses have smaller radiative envelopes which causes their
eigenfrequencies to be larger and more widely spaced than those of a
lower-mass ZAMS star. For a more extended discussion of the relative
differences between $k_2$ and $k_{\rm 2,dyn}$ for ZAMS stars, we refer
to Smeyers \& Willems (2001).

The extent of the deviations caused by the compressibility of the
stellar fluid also depends on the evolutionary stage of the star
(Willems \& Claret 2002). The dependency is primarily through the
evolution of the radius and the associated change of the star's
dynamical time scale. In particular, as the star evolves on the main
sequence, the radius and the dynamical time scale increase so that in
relative terms the orbital period becomes shorter in comparison to the
star's dynamical time scale. Correspondingly, the forcing frequencies
become larger when expressed in units of the inverse of the star's
dynamical time scale.  The deviations due to the compressibility of
the stellar fluid are therefore larger for a model with a larger
radius.  The evolution of the star furthermore also affects the
deviations caused by the resonances of dynamic tides with free
oscillation modes of the component stars. In particular, the effects
of the resonances tend to be larger for stars near the end of
core-hydrogen burning than for stars on the zero-age main
sequence. This effect is again related to the size of the radiative
envelope, which increases which increasing age of the star on the main
sequence.

\section{Tides in compact object binaries}

A sometimes under appreciated aspect of the study of tides in close
binaries is its contribution to understanding the {\em past} formation
and evolution of binaries and their component stars. This is
particularly true for binaries containing compact objects such as
neutron stars and black holes. These objects are known to be formed in
violent supernova explosions or core collapse events during which a
large fraction of the system mass is lost and asymmetries in the
stellar interior can impart kick velocities of several hundreds of
km/s to the compact object at birth.  The mass loss and supernova
kicks can either expand or shrink the orbit and almost always induce a
non-zero eccentricity. X-ray binaries in which a neutron star or black
hole accretes matter from a Roche-lobe filling companion, however,
typically have nearly circular orbits. Tidal circularization must
therefore have been active sometime between the formation of the
compact object and the onset of the mass-transfer phase (unless the
mass-transfer process itself provides an efficient circularization
mechanism). A detailed understanding of tidal effects is then
essential if one wishes to reconstruct the evolutionary history of the
binary to learn something about the formation of the compact object
and its progenitor\footnote{Note that such a reconstruction is only
possible if the present-day orbital eccentricity is not identically
equal to zero.}. Willems et al. (2005), e.g., showed that for
GRO\,J1655-40, a more accurate quantitative knowledge of tidal effects
would tighten the constraints on the progenitor of the central black
hole (an OB-star stripped from its hydrogen-rich envelope) by as much
as 30\%.

Another promising future application involving tidal effects in compact
object binaries is the study of gravitational waves emitted by double
white dwarfs. These objects originate from binaries initially
consisting of two B stars and are expected to be the single most
abundant astrophysical sources guaranteed to be detectable by the
Laser Interferometer Space Antenna ({\em LISA}) scheduled to launch in
2015. Double white dwarfs with orbital periods longer than $\simeq
1000$\,s are expected to emit gravitational waves with a constant
amplitude and a constant frequency equal to twice the orbital
frequency. For shorter-period systems, the emission of gravitational
waves drains the orbital energy, causing the white dwarfs to spiral in
towards each other. So far, all predictions for this inspiral are
based on the assumption that both white dwarfs can be treated as point
masses. However, as the orbit decays, finite size effects become
increasingly important, especially when the orbital frequency sweeps
through the spectrum of eigenmode frequencies, and tidal effects
inevitably contribute to the orbital inspiral. Since the detection of
gravitational waves by laser interferometers is done by template
matching techniques, the inclusion of these effects in predicted
gravitational wave signals is imperative to the successful detection
of double white dwarfs by {\em LISA}. In this way, successful
detections will also yield a wealth of information on the physics of
white dwarfs, and thus the cores of their B-star progenitors, which
may prove inaccessible through standard electromagnetical windows on
the Universe used by, e.g., asteroseismologists.

\section{Future prospects and concluding remarks}

Tides in close binaries give rise to a wide range of dynamical effects
with vastly different orders of magnitude. Their physical origin can
be as simple as the differential gravitational attraction giving rise
to the tidal bulge, and as complicated as the convective and radiative
damping mechanisms responsible for circularizing the orbit and
synchronizing the binary components with their orbital motion. A brief
summary of the driving mechanism behind different tidal effects and
the associated orders of magnitude is given in Table~\ref{tides}.

\begin{table}
\caption{Summary of the origin and order of magnitude of tidal
  effects in close binaries. } 
\label{tides} 
\smallskip
\begin{center}
\begin{tabular}{lcc}
\tableline
\noalign{\smallskip}
Tidal Effect & Origin & Order of Magnitude \\
\noalign{\smallskip}
\tableline
\noalign{\smallskip}
tidal bulge     & differential gravitational attraction & $(R_1/a)^3$ \\ 
apsidal motion  & perturbation of the gravitational field & $(R_1/a)^5$ \\
synchronization & energy dissipation/tidal torque & $(R_1/a)^6$ \\
orbital decay   & energy dissipation/tidal torque & $(R_1/a)^8$ \\
circularization & energy dissipation/tidal torque & $(R_1/a)^8$ \\
\noalign{\smallskip}
\tableline
\end{tabular}
\end{center}
\end{table}

The variety in the order of magnitude and physical background of the
effects listed in Table~\ref{tides} makes tides a potentially very
versatile probe of different regimes of stellar physics. The main
difficulty in using tides as asteroseismological probes is currently
posed by the quality of the observations. Firstly, the steep
dependence of tidal effects on the ratio $R_1/a$ demands very accurate
stellar radii and orbital separations in order to get reliable orders 
of magnitude. Secondly, in binaries that are close enough for
dynamical tides to be important, the dependence of the tidal forcing
frequencies on the rotational angular velocity requires accurate
determinations of the stellar rotation rates in order to properly
account for resonant excitations of free oscillation modes.

Any hopes of realizing the required accuracies in the near future most
likely rest on vigorous studies of eclipsing binaries in which the
eclipse geometry allows precise determinations of the stellar radii
and the strongly constrained orbital inclination helps pinning down
the components' rotation rates. In view of the spectacular number of
ongoing and upcoming large-scale photometric variability and exoplanet
transit surveys, the odds of finding eclipsing binaries with
fortuitous system parameters are increasingly favorable though. The
next decade therefore seems to be an extremely promising era for the
exploitation of tides as probes to study the formation and evolution
of binaries with component stars across the entire Hertzsprung-Russell
diagram.

\acknowledgments I would like to express my sincere thanks to the
organizers for the opportunity to present this talk at the meeting and
for the financial support allowing me to travel to Sapporo. I would
also like to thank Vicky Kalogera for continuous support through a
David and Lucile Packard Foundation Fellowship in Science and
Engineering grant and NASA ATP grant NAG5-13236.

\noindent
{\bf S. Owocki:} Could you comment on the role of the tidal forcing
effects on cases where the star is rapidly rotating, i.e.\ near
critical rotation? \\

\noindent
{\bf B. Willems:} The study of tides in rapidly rotating stars suffers
the same difficulties as the study of free oscillations in rapidly
rotating stars. Currently there is no theory available to deal with
either of them. Once a theory describing the effects of rapid rotation
on free oscillation modes becomes available, a generalization to tides
should be fairly straightforward. \\

\noindent
{\bf A. T. Okazaki:} In a Keplerian circumstellar disk in a highly
eccentric system, the particles in the outer parts of the disk have
the same frequency as the instantaneous orbital frequency of the
companion. Do resonant dynamic tides work on these particles in the
outer disk? \\

\noindent
{\bf B. Willems:} Yes, although the problem evidently has a different
base geometry than resonances in a spherically symmetric equilibrium
star. Lubow (1981), e.g., has shown that resonant accretion disk tides can
generate horizontally propagating waves which contribute to the
transfer of disk angular momentum to orbital angular momentum. \\

\noindent
{\bf J. Bjorkman:} For a wide, highly eccentric binary with a close
periastron passage, there will be a characteristic time scale for the
tidal interaction.  Is it possible that such an interaction will
resonantly excite stellar pulsations for a few e-foldings of their
growth? If so, can you describe what conditions would be favorable for
this? Such a mechanism might act as a trigger for mass ejections from
a near-critically rotating star. \\

\noindent
{\bf B. Willems:} As long as the orbital period is short in comparison
to the mode damping or e-folding time, the oscillations will be
sustained and ``re-excited'' during each successive periastron
passage. The conditions for resonant excitation become increasingly
favorable with decreasing orbital period and increasing orbital
eccentricity (i.e. with decreasing periastron passage time). \\ 

\noindent
{\bf G. Meynet:} The deformation of the star due to tidal forces might
induce some kind of meridional circulation. Do you know if this
process is important? Could it induce some kind of mixing of the
elements? \\

\noindent
{\bf B. Willems:} I am not aware of any work looking into tidally
induced meridional circulation. It is conceivable, however, that the
tidal distortion creates a temperature difference between the poles
and the equator similar to that due to rotational flattening. Whether
or not any resulting meridional circulation penetrates deep enough
into the star to mix the elements remains to be seen. \\

\end{document}